# Metastable $MnBi_2Te_4$ enabled by magnetic-field-assisted synthesis

Abhinna Rajbanshi[1], G. M. Zills[2], A. M. Donald[3], Daniel Duong[1], David Graf[4], James Hamlin[3], M. W. Meisel[3,5], I. Vekhter[2], Williams A. Shelton[2,6], and Rongying Jin[1,*]
(Corresponding author: Rongying Jin, Email: rjin@mailbox.sc.edu)

[1]*SmartState Center for Experimental Nanoscale Physics, Department of Physics and Astronomy, University of South Carolina, Columbia, SC 29208, USA*

[2]*Department of Physics and Astronomy, Louisiana State University, Baton Rouge, LA 70803, USA*

[3]*Department of Physics, University of Florida, Gainesville, FL 32611-8440, USA*

[4]*National High Magnetic Field Laboratory, Tallahassee, FL, 32310, USA*

[5]*National High Magnetic Field Laboratory High B/T Facility, University of Florida, Gainesville, FL 32611-8440, USA*

[6]*Center for Computation & Technology, Louisiana State University, LA 70803, USA*

## ABSTRACT

Magnetic topological insulators provide a unique platform to explore the interplay between magnetism and topology. $MnBi_2Te_4$, known for its A-type antiferromagnetic (AFM) ground state, undergoes a striking transformation when single crystals are grown in an applied magnetic field. Despite retaining the same crystal structure, field-grown $MnBi_2Te_4$ exhibits a ferromagnetic (FM) ground state with a Curie temperature of ~12.5 K, confirmed by magnetization, magnetic torque, electrical resistivity, and specific heat measurements. First-principles calculations support these findings, revealing that magnetic-field-assisted synthesis can effectively reconfigure the ground-state spin order and thereby modify the material's electronic properties, as reflected in the de Haas-van Alphen oscillation seen in the magnetic torque.

## 1. INTRODUCTION

Magnetic topological materials, which uniquely combine nontrivial electronic topology with the intrinsic magnetic order, represent a rapidly advancing frontier in materials physics. These systems host a wide variety of exotic quantum states such as the quantum anomalous Hall effect (QAHE), axion electrodynamics, chiral Majorana modes, higher-order topological phases [1-4] that challenge conventional understanding and open new avenues for exploring fundamental physical principles. The rich interplay between band topology and magnetism not only deepens our knowledge of how charge and spin degrees of freedom can interwine but also establishs this field as one of the most exciting and intriguing areas of materials research [4-6]. Beyond their fundamental scientific significance, the realization and control of these quantum phenomena provide an exceptional platform for envisioning and engineering next-generation technologies. In particular, they offer promising pathways toward innovative spintronic architectures, topological quantum computation schemes, and other device concepts that leverage the robustness and tunability of topologically protected states [7-8].

A central challenge is the in-depth understanding and precise controll of magnetic ordering in these materials, since the nature of the magnetic configuration plays a decisive role in dictating the emergence, robustness, and tunability of the associated topological phases. Subtle variations in magnetic anisotropy, exchange interactions, or spin textures can significantly alter the electronic band topology, making it essential to develop strategies—both theoretical and experimental—that enable reliable manipulation of magnetic states. Achieving such control is critical not only for uncovering the fundamental mechanisms linking magnetism and topology but also for realizing stable, functional topological phases suitable for technological applications.

Layered van der Waals compound $MnBi_2Te_4$ (Figure 1(a)) is a well-known intrinsic magnetic topological insulator, integrating the nontrivial electronic band topology inherited from the $Bi_2Te_3$ building block and the A-type antiferromagnetic (AFM) ordering between MnTe septuple layers [9–19]. Owing to the weak interlayer coupling between the $Bi_2Te_3$ and MnTe substructures, the magnetic ground state of $MnBi_2Te_4$ is susecptibile to external perturbations. For example, both the magnetic field and hydrostatic pressure can continously tune the system from its native AFM configuration [19–27] to a ferromagnetic (FM) state [11,25,28–36]. Under sufficiently strong magnetic fields, the Chern insulator phase emerges, exhibiting a robust zero Hall-resistance plateau [31]. However, these field-induced phases vanish once the external stimulus is removed. A recent study suggests that the energy difference between the A-type AFM state and the compensated FM state for $MnBi_2Te_4$ is only 8.694 meV, equivalent to ≈ 100 K [28]. This small energy difference implies that even moderate thermal or magnetic perturbations can significantly reshape the energy landscape. Indeed, magnetic-field annealing of $MnBi_2Te_4$ at 600 K has been shown to destabilize the A-type AFM ground state and drive the system into a superparamagnetic state [28]. Such remarkable magnetic tunability demonstrates that $MnBi_2Te_4$ provides an effective platform for accessing metastable magnetic states, enabling systematic exploration of their associated quantum phases and emergent physical properties.

In this article, we report the synthesis of single crystals of $MnBi_2Te_4$ in an external magnetic field, and the characterization of its ground state structure and physical properties via x-ray diffraction, magnetic torque, magnetization, electrical resistivity, and specific heat measurements. While the crystal symmetry remains unchanged, field-grown $MnBi_2Te_4$ exhibits a FM ground state with the Curie temperature ($T_C \approx 12.5$ K), which is different from the AFM state for zero-field-grown $MnBi_2Te_4$. This difference leads to different temperature and magnetic field dependence in the magnetization, magnetotransport, and magnetic torque. Quantitative analysis of the data at low temperatures suggests new electronic properties in the field-grown $MnBi_2Te_4$.

## 2. EXPERIMENTAL DETAILS

Zero-field-grown single crystals of $MnBi_2Te_4$ were obtained following the procedure described in Ref. [28]. Field-grown $MnBi_2Te_4$ single crystals were synthesized via the self-flux method in the continuous presence of a 9 T magnetic field using the $B\times T$ instrument of the National High Magnetic Field Laboratory (NHMFL), Gainsville, Florida [37]. First, $Bi_2Te_3$ and MnTe ingots were mixed in a ratio of 1 : 1 and placed into an evacuated quartz tube. The sealed quartz tube was placed in a furnace with a 9 T provided by a superconducting magnet, slowly heated from ambient to 700 ºC in 14 hours, and then kept at constant temperature for 10 h. The sample was subsequently cooled to 590 ºC in 110 h, followed by quench to water. During the entire crystal growth process, the magnetic field of 9 T was present. The inset of Figure 1(b) shows a field-grown $MnBi_2Te_4$ single crystal with the typical size of ~1.5 mm × 1.0 mm × 0.5 mm.

The crystal structure of field-grown $MnBi_2Te_4$ was verified by x-ray diffraction (XRD), Rigaku Ultima IV, with Cu K$\alpha$ radiation ($\lambda_{K\alpha1}$ = 1.5406 Å and $\lambda_{K\alpha2}$ = 1.5444 Å). The magnetization was measured using vibrating sample magnetometer (VSM) option of DynaCool (*Quantum Design*, PPMS – 14 T). Electrical resistivity (up to 14 T) and Hall effect measurements were also performed with the PPMS using the four-probe method. Specific heat data were obtained with a 2.8 mg $MnBi_2Te_4$ single crystal with the PPMS employing the relaxation method. Low temperature (down to 0.4 K) and high magnetic field (± 35 T) magnetoresistance (MR) and magnetic torque measurements were conducted at the NHMFL (Tallahassee, Florida). The magnetic torque data were obtained with a tiny single crystal, which was secured to a cantilever with grease, and the magnetic field was swept between -35 T and +35 T at a rate of 3 T/min at a fixed temperature.

We performed first-principles density functional theory (DFT) calculations. The structural, electronic, and magnetic properties were obtained using the Vienna Ab Initio Simulation

Package (VASP) [38-40]. We used projector augmented wave (PAW) pseudopotentials [41] for Mn (3s, 3p, 3d, 4s), Bi (5d, 6s, 6p), and Te (5s, 5p). A plane wave energy cutoff of 720 eV and a 12 × 12 × 8 Monkhorst – Pack k-space integration method was used. Relativity was employed in all calculations, including those involving non-collinear magnetic structures. In addition, the van der Waals (vdW) interaction was included via the non-local vdW-DF optB88 functional [42] as implemented in VASP.

## 2. RESULTS AND DISCUSSION

Figure 1(b) shows the XRD patterns obtained from a flat surface of a single crystal for zero-field-grown (black) and field-grown (red) $MnBi_2Te_4$, respectively. In each case, all XRD peaks can be indexed with (0 0 *l*) (*l* = 3, 6, 9,…) for the space group R-3m, indicating the crystal symmetry of the field-grown crystal is the same as the zero-field-grown $MnBi_2Te_4$, albeit with subtle variations of the lattice parameters. Note the red peaks are slightly shifted compared to the peak positions for the zero-field-grown crystal as shown in Figure 1(c) for the (0 0 6) peak. Correspondingly, the lattice parameter $c$ is ≈ 40.8619 Å for the zero-field-grown crystal and ≈ 40.7881 Å for the field-grown one (a change of ~0.2%). On the other hand, the (0 0 6) peak reveals similar full width half maximum for each case, suggesting similar crystalline quality.

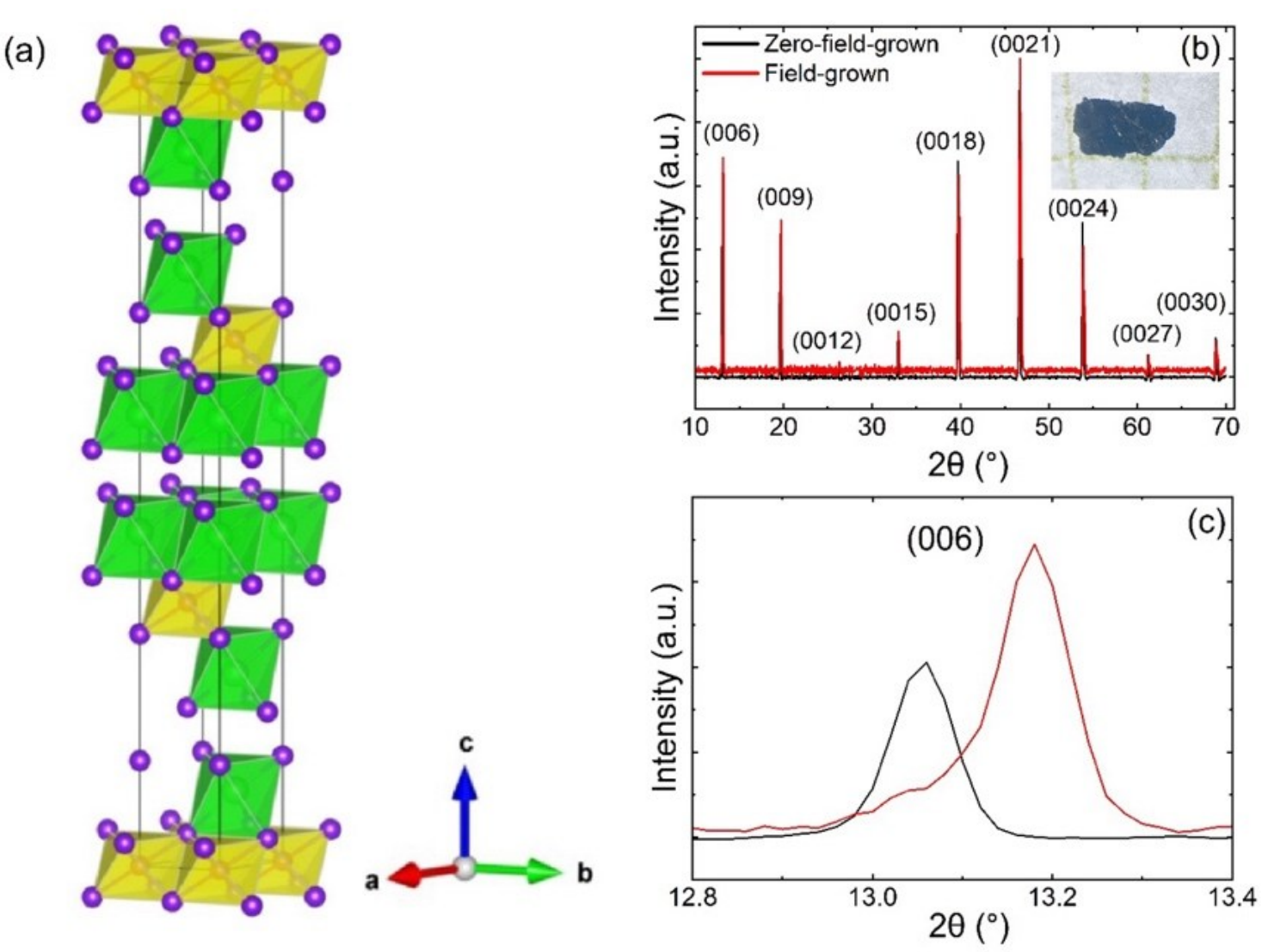

FIG. 1. (a) Crystalline structure of $MnBi_2Te_4$. (b) X-ray diffraction patterns for zero-field-grown (black) and field-grown $MnBi_2Te_4$ (red) single crystals, respectively. The inset shows a photo of a field-grown $MnBi_2Te_4$ single crystal. (c) Zoom-in around the (0 0 6) peak.

Figure 2(a) presents the temperature dependence of the magnetic susceptibility of field-grown $MnBi_2Te_4$ obtained by applying the magnetic field H = 1000 Oe along the *c* axis ($\chi_c$) and *ab* plane ($\chi_{ab}$) between 2 and 400 K. With decreasing temperature, both $\chi_c$ and $\chi_{ab}$ slowly increase with a sharper rise below 15 K, as enlarged in the inset of Figure 2(a). There is little difference between the field cooling (FC) and zero field cooling (ZFC) magnetic susceptibility along the *c* axis and *ab* plane until reaching $T_x \sim 4.5$ K. To explore the magnetic interactions at high temperatures, nominally over the range of 30 K to 300 K, both $\chi_c$ and $\chi_{ab}$ were fit using the Curie-Weiss (CW) formula $\chi(T) = \chi_0 + \frac{C}{T-\theta_{CW}}$, where $\chi_0$ is the temperature independent magnetic susceptibility, C is the Curie constant, and $\theta_{CW}$ is the Curie-Weiss temperature, and the results are listed in Table I. The positive $\theta_{CW}^{c}$ and $\theta_{CW}^{ab}$ imply dominant ferromangetic interaction along both the *c* axis and *ab* plane. The effective magnetic moments ($\mu_{eff} = \sqrt{8C}\mu_B$) obtained from the Curie constant are $\mu_{eff}^{c} \sim 2.25\mu_B$ and $\mu_{eff}^{ab} \sim 2.21\mu_B$. To check the fitting quality, the magnetic susceptibility is replotted as $\frac{1}{\chi-\chi_0}$ versus T for both $\chi_c$ and $\chi_{ab}$, and the fitting results represented by solid lines are given in Figure 2(b). Note that both $\chi_c$ and $\chi_{ab}$ are well fitted by the Curie-Weiss formula between 30 K and 300 K.

Compared to zero-field-grown $MnBi_2Te_4$ [10,11,28,43-45], the magnetic susceptibility for the field-grown $MnBi_2Te_4$ provides several dramatically different features: (1) positive $\theta_{CW}^{c}$, (2) small $\mu_{eff}^{c}$ and $\mu_{eff}^{ab}$, and (3) increases in $\chi_c$ and $\chi_{ab}$ below 25 K. For easy observation, we plot $d\chi_c/dT$ and $d\chi_{ab}/dT$ in Figure 2(c), which show the maximum change at $T_C \approx 12.5$ K and $T_x \sim 4.5$ K. Below $T_x$, FC and ZFC susceptibilities are distinguished. All of these outcomes indicate the nature of magnetic interaction in the field-grown $MnBi_2Te_4$ is different compared to samples grown in zero applied field.

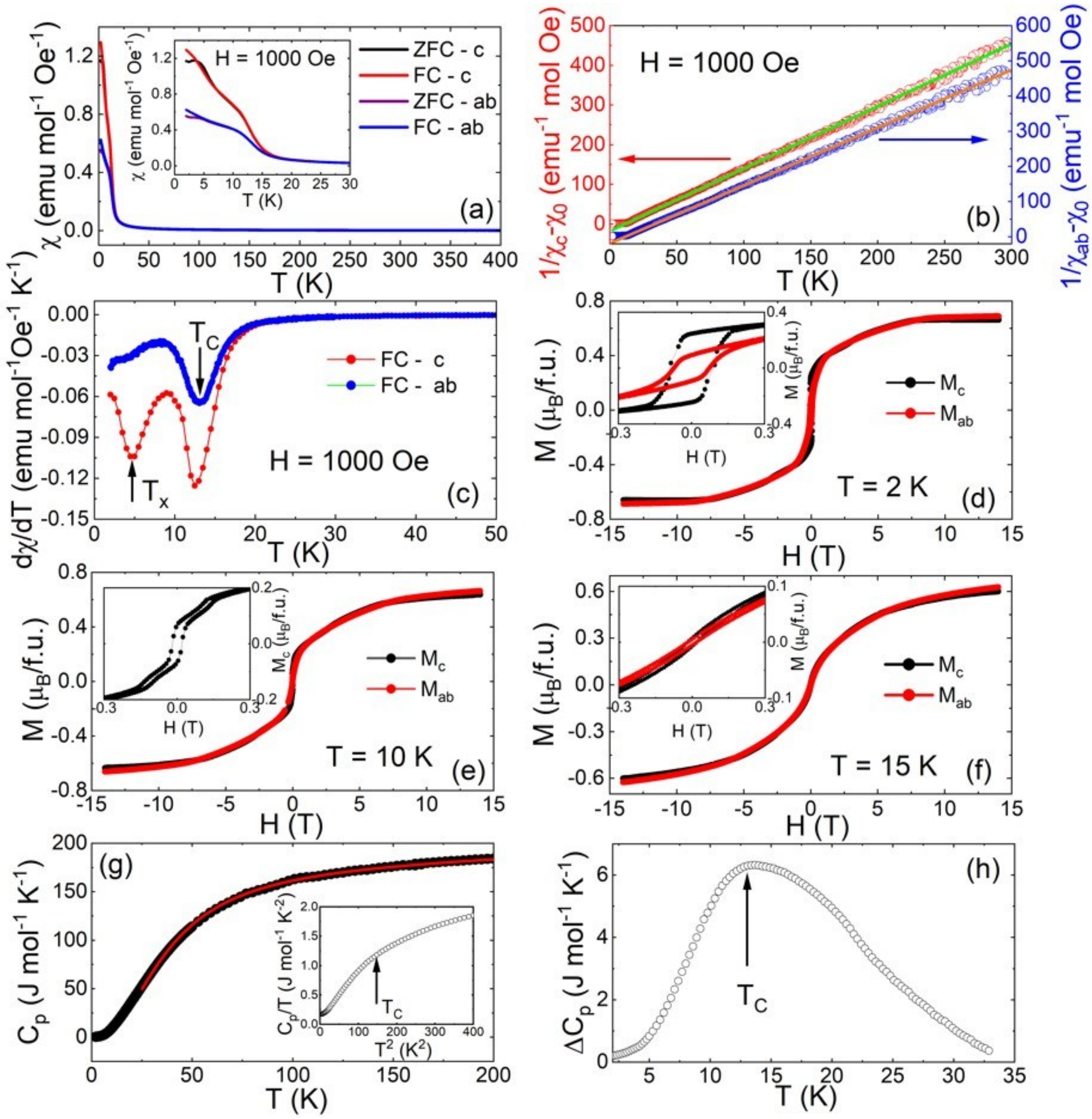

FIG. 2. (a) Temperature dependence of the magnetic susceptibility of field-grown $MnBi_2Te_4$ with the magnetic field H = 1000 Oe applied along *c* axis ($\chi_c$) and within *ab* plane Type equation here. ($\chi_{ab}$). (b) Comparison of experimental $\chi_c$ and $\chi_{ab}$ with the Curie-Weiss fit results represented by solid curves. (c) Derivative of the magnetic susceptibility, indicating $T_x$ and $T_C$. (d-f) Field dependence of the magnetization at 2 K (d), 10 K (e) and 15 K (f). Insets are zoomed-in magnetization for both the *c* and *ab* directions. (g) Temperature dependence of the specific heat. The red line represents the fitting of data with Eq. (1). Inset: low-temperature specific heat plotted as $C_p/T$ vs. $T^2$. (h) Temperature dependence of the specific heat after subtracting electronic and phononic contributions ($\Delta C_p$).

| | $\chi_0$ (emu/mol-Oe) | $\theta_{CW}$ (K) | C (emu-K/mol-Oe) | $\mu_{eff}$ ($\mu_B$) |
|---|---|---|---|---|
| H // *ab* | $-2.49 \times 10^{-4}$ | 11.6 | 0.611 | 2.21 |
| H // c | $-7.08 \times 10^{-4}$ | 11.7 | 0.633 | 2.25 |

TABLE I: Parameters obtained by fitting experimental $\chi_{ab}$ and $\chi_c$ between 30 K and 300 K to the Curie-Weiss formula (see text).

To characterize the magnetic behavior at low temperatures, the isothermal magnetization of field-grown $MnBi_2Te_4$ was measured. Shown in Figures 2(d-f) is the magnetic field dependence of the magnetization for T = 2 K, 10 K and 15 K, respectively. Both $M_c$ and $M_{ab}$ exhibit strong field dependence with sharp changes at low fields and signs of saturation at high fields. At T = 2 K, both $M_c$ and $M_{ab}$ exhibit a hysteresis loop as shown in the inset of Figure 2(d). The remanent magnetization ($M_r^c \sim 0.25\mu_B$/f.u.) and coercive field ($\mu_0 H_{co}^c \sim 0.08$ T) for the *c* direction are larger than those ($M_r^{ab} \sim 0.09\mu_B$/f.u. and $\mu_0 H_{co}^{ab} \sim 0.07$ T) for the *ab* plane, thereby indicating anisotropy with the magnetic easy axis along the *c* direction. As can be seen in the inset of Figure 2(e), there is also a magnetic hysteresis loop in $M_c$ at T = 10 K, revealing step like feature. At T = 15 K, the hysteresis loop can no longer be identified as shown in the inset of Figure 2(f). These results clearly indicate field-grown $MnBi_2Te_4$ possesses a ferromagnetic ground state with the Curie temperature $T_C \approx 12.5$ K.

The effective magnetic moment (see Table I) estimated from high-temperature susceptibilities suggests that the total spin S = ½ for Mn, i.e., in a state with a low spin for field-grown $MnBi_2Te_4$. Given that zero-field-grown $MnBi_2Te_4$ has total spin of S = 5/2 for Mn [44], it becomes clear that the presence of the magnetic field during the growth of $MnBi_2Te_4$ lifts the degenercy between the $t_{2g}$ and $e_g$ orbitals, leading to the placement of 5 electrons in $t_{2g}$ orbitals. This is striking as $Mn^{2+}$ prefers in the high spin state with half filling. Mn-Te forms a octahedral type of structure with trigonal elongation that reduces the splitting between $t_{2g}$ and $e_g$ orbitals. Even with spin orbit coupling coming from Bi and Te, the crysal field splitting does not compete with Hund's coupling. It may be possible to obtain a low spin state via high compression that would act opposite to the trigonal elongation. While the field-grown $MnBi_2Te_4$ exhibits a slightly shorter lattice parameter *c* (see Fig. 1), it is difficult to drive the system into a low spin configuration. Another possibility is that defects change the Mn valence producing low-spin state type behavior, such as Te vacancies, antisities (Mn replacing Bi), interstitials and stacking faults. These defects can locally compress the octahedron, i.e., reducing the elongation.

Nevertheless, at T = 2 K, M(H) saturates at ~ $0.65\mu_B$/f.u. for both the *ab* plane and *c* direction (see Figure 2(d)) suggests the compensated ferromagnetism with the up-down-up configuration. Based on DFT calculations [28], such a magnetic state is metastable with the higher energy (8.694 meV) than the zero-field-grown $MnBi_2Te_4$. Considering $Mn^{2+}$ 3*d* states hybridizing with Te *p* states, non-Heisenberg terms (e.g. biquadraic exchange) can become important (compared with the Heisenberg term). This and strong easy-axis anisotropy can lead to a compensated FM state. The step-like feature seen in $M_c$ between $T_x$ and $T_C$ (inset of Figure 2(e)) suggests the alignment of domains under different fields. Altough there is no hysterisis, the strong nonlinearity of M(H) at 15 K (Fig. 2(f)) implies that there is strong FM fluctuation above $T_C$.

With the small saturation moment at 2 K (Figure 2(d)), it is important to investigate the associated entropy change across $T_C$ and $T_x$. Figure 2(g) displays the temperature dependence of the specific heat of field-grown $MnBi_2Te_4$ between 2 K and 200 K, which shows no anomaly at $T_C$ nor at $T_x$. By plotting the low-temperature data as $C_p/T$ vs $T^2$ in the inset of Figure 2(g), a noticable, broad slope change is detected across $T_C$. Below $T_C$, $C_p/T$ vs. $T^2$ is nonlinear, implying $C_p$ cannot be simply explained by considering only electron ($\gamma T$) and phonon ($\beta T^3$) contributions, where $\gamma$ is Sommerfeld electronic specific heat coefficient and $\beta$ is a constant. Nevertheless, the specific heat above 30 K can be well fitted to the Debye-Einstein model [46]

$$C_P = \gamma T + 9Rn_D\left(\frac{T}{\theta_D}\right)^3\int_0^{\frac{\theta_D}{T}}\frac{x^4e^x dx}{(e^x-1)^2} + 3Rn_E\left(\frac{\theta_E}{T}\right)^2\frac{e^{\frac{\theta_E}{T}}}{\left(e^{\frac{\theta_E}{T}}-1\right)^2}. \quad (1)$$

Here, the second and third terms represent Debye and Einstein descriptions of phonon contribution to the specific heat with R being the gas constant, $n_D$ and $n_E$ the numbers of the Debye and Einstein modes, respectively. For $MnBi_2Te_4$, $n_D = 3$ and $n_E = 2$, making $n_D + n_E = 5$, where $\theta_D$ is the Debye temperature and $\theta_E$ the Einstein's temperature. The red curve in Figure 2(g) represents the fit using Eq. (1) with $\gamma = 71 \pm 1$ mJ mol$^{-1}$ K$^{-2}$, $\theta_D = 180.0 \pm 0.4$ K, and $\theta_E = 73.2 \pm 0.4$ K. With these parameters, the additional specific heat, $\Delta C_p = C_p$(measured)-$C_p$(Eq.(1)), at low temperatures can be subtracted, which is shown in Figure 2(h). Note that $\Delta C_p$ becomes nonzero below ~ 30 K and reveals a broad peak at $T_C$, implying entropy removal by the ferromagnetic transition. Similar entropy changes were observed in $MnSb_2Te_4$ with with nonzero magnetic specific heat below $2T_C$ ~ 50 K and reaches the maximum at $T_C$ [47].

Given the finite electronic specific heat $\gamma$ value, the system is expected to be conductive. Figure 3(a) shows the temperature dependence of the in-plane ($\rho_{ab}$) and out-of-plane ($\rho_c$) resistivity between 2 K and 300 K. Both $\rho_{ab}$ and $\rho_c$ monotonically decrease with decreasing temperature, revealing a kink at $T_C$, which indicates reduced spin scattering below $T_C$ in both $\rho_{ab}$ and $\rho_c$. Applying H // *c*, the kink in $\rho_{ab}$ is quickly smeared out. For H > 1 T, $\rho_{ab}$ gradually resumes $T^2$ dependence. As shown in Figure 3(b), the $T^2$ dependence of $\rho_{ab}$ at 12 T extends to 30 K, implying the recovery of the Fermi-liquid behavior in high fields.

To understand the evaluation of the resistivity under the magnetic field, the field dependence of the in-plane magnetoresistance ($MR_{ab} = \frac{\rho_{ab}(H)-\rho_{ab}(H=0)}{\rho_{ab}(H=0)} \times 100\%$) was measured between -14 T and 14 T by applying H // *c* , and the MR data between 2 K and 50 K are shown in Figure 3(c). An expanded view of the low-field region is shown in the inset, where $MR_{ab}$ is negative between -2 T and 2 T for $T \leq 15$ K. The negative $MR_{ab}$ is attributed to the suppression of spin scattering due to spin alignment by the magnetic field. Above ~2 T, the conventional orbital $MR_{ab}$ exceeds the negative contribution, leading to a sign change of $MR_{ab}$. Above 15 K, $MR_{ab}$ is positive at all applied fields, indicating that the spin-induced $MR_{ab}$ is negeligible away from $T_C$. More interestingly, $MR_{ab}$ tends to increase linearly with H without sign of saturation at all measured temperatures (2 K ≤ T ≤ 50 K). To confirm the linear behavior, additional studies were conducted in higher fields and at lower temperatures. Figure 3(d) shows the field dependence of $MR_{ab}$ of field-grown $MnBi_2Te_4$ at T = 0.4 K for various H directions between the *c* and *ab* plane. Two features can be seen: (1) $MR_{ab}$ exhibits linear H dependence in all applied field directions and amplitude (up to 35 T) and (2) $MR_{ab}$ at a given field varies monotonically with θ with the largest amplitude at H // *c* (105% at 35 T and θ = 0°) and lowerest at H // *ab* (27% at 35 T and θ = 90°). Such magneto anisotropy is likely related to the electronic structure of field-grown $MnBi_2Te_4$.

For zero-field-grown $MnBi_2Te_4$, the electronic properties are governed by the surface states because the bulk is gapped [48-49]. Both angle-resolved photoemission spectroscopy (ARPES) and Hall effect measurements indicate that the dominant carriers are electrons [50]. Figure 3(e) presents the magnetic field dependence of the Hall resisitivity $\rho_{xy}$ of field-grown $MnBi_2Te_4$ between 2 K and 30 K. Different from the zero-field-grown case [50], $\rho_{xy}$ for field-grown $MnBi_2Te_4$ exhibits little temperature dependence. Furthermore, the field dependence of $\rho_{xy}$ is nearly linear up to 14 T with the positive slope $R_H$ ~ 0.00263 $cm^3/C$, which gives the carrier concentration ~$2.4\times10^{21}$ $cm^{-3}$ through the Drude model. This result strongly suggests that the electronic structure of field-grown $MnBi_2Te_4$ is different from that of the zero-field-grown case. Figure 3(f) presents the field dependence of $\rho_{xy}$ at 2 K for zero-field-grown (green), field-annealed (red), and field-grown $MnBi_2Te_4$ (black). A significant difference in $\rho_{xy}$ among the three cases reflects the magnetic field effect in magnetic materials synthesis and processing. In particular, the sign change of the slope from negative for the zero-field-grown case to positive for the field-grown case clearly indicates the modification of the electronic properties.

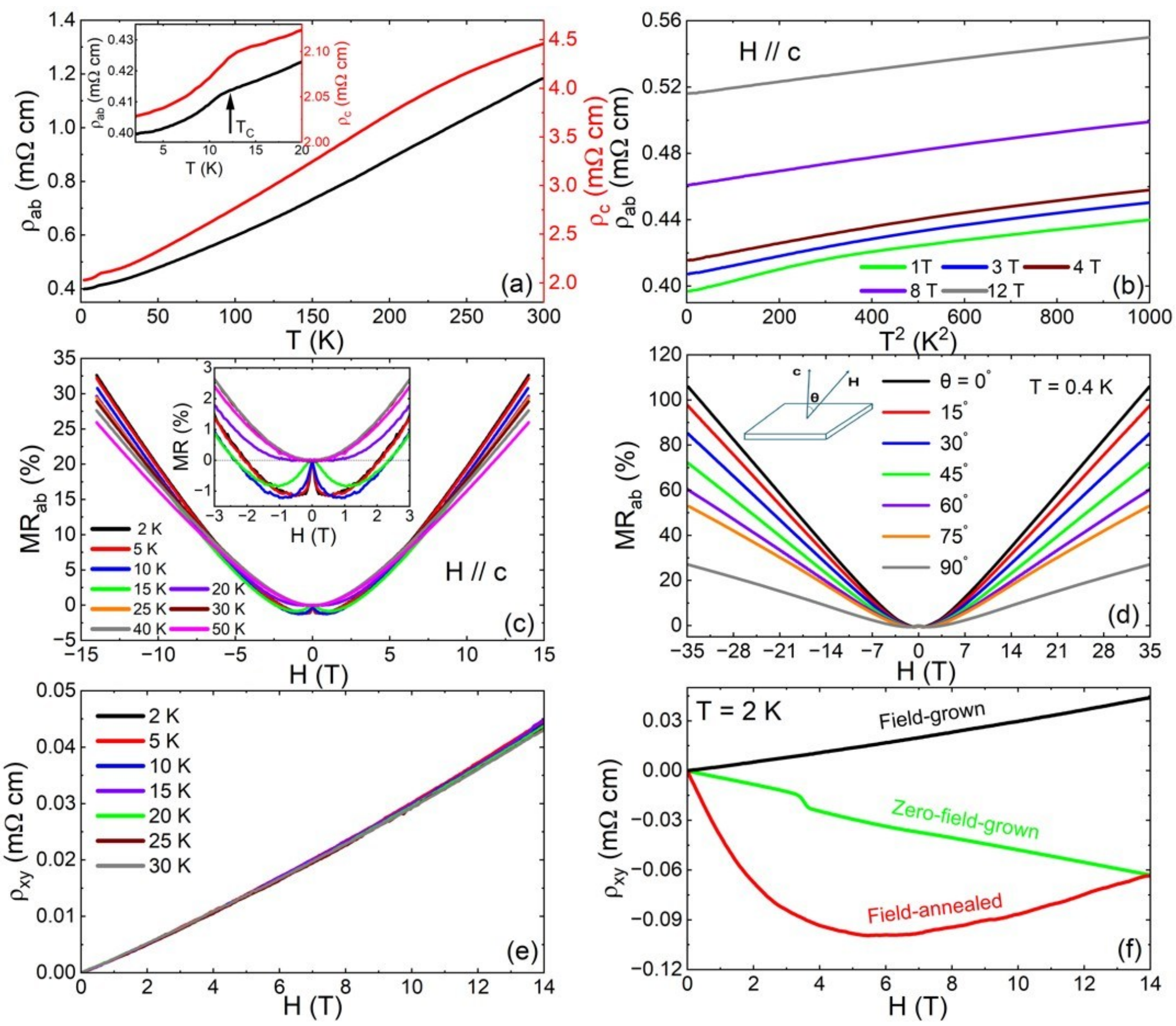

FIG. 3. (a) Temperature dependence of the in-plane ($\rho_{ab}$) and out-of-plane ($\rho_c$) of field-grown $MnBi_2Te_4$. (b) Low-temperature $\rho_{ab}$ at different magnetic fields applied along the *c* axis plotted as $\rho_{ab}$ vs. $T^2$. (c) Magnetic field dependence of $MR_{ab}$ with H // *c* (θ=0°) at different temperatures. Inset: MR between -3 T and 3 T. (d) Field dependence of MR at T = 0.4 K with the magnetic field direction varying between the *c* and *ab* directions. (e) Magnetic field dependence of the Hall resistivity ($\rho_{xy}$) at different temperatures with H // *c*. (f) Comparison of $\rho_{xy}$ for zero-field-grown (green), field-annealed (red) and field-grown (black) $MnBi_2Te_4$ at T = 2 K.

The difference between field-grown and zero-field-grown $MnBi_2Te_4$ is further illustrated in Figure 4. Figures 4(a-b) display the field dependence of the magnetic torque (τ) of zero-field-grown and field-grown $MnBi_2Te_4$ with the field applied along the *c* axis (θ = 0°) and *ab* plane (θ = 90°), respectively. In both θ = 0° and 90°, τ for field-grown $MnBi_2Te_4$ exhibits nearly opposite field dependence to that for zero-field-grown $MnBi_2Te_4$, reflecting the different spin configurations between them. While other work has established the magnetic alignment along the *c* direction in zero-field-grown $MnBi_2Te_4$ is antiferromagnetic [19-27], field-grown $MnBi_2Te_4$ has a net magnetic moment along the *c* direction resulting in a sharp initial increase of τ. Given that $\vec{\tau} = \vec{M} \times \vec{H}$ and M varies monotonically with H (Figs. 2(d-f)) until saturation around 60 T [44], the field dependence of τ is soley determined by the angle between $\vec{M}$ and $\vec{H}$, i.e., $\tau \propto \sin(\widehat{\vec{M}\,\vec{H}})$. Following the characterization for zero-field-grown $MnBi_2Te_4$ in Ref. [44], there are three characteristic fields $H_{SF}$, $H_1$, and $H_2$, describing the onset of spin-flop (SF), spin flip of the normal Mn moment, and spin flip of the antisite Mn moment, respectively. For convinence, these characteristic fields are labeled in Figure 4(b). Figures 4(c-d) show the field dependence of τ at $0° < \theta < 90°$ for field-grown and zero-field-grown $MnBi_2Te_4$, respectively. Note both samples possess nearly identical field and angle dependence of τ, impling that the spin structure for field-grown $MnBi_2Te_4$ cannot be purely ferromagnetic but instead a compensated FM, as discussed above. Interestingly, the characteristic fields, $H_{SF}$, $H_1$ and $H_2$, increase monotonically upon the rotation of the applied field from the *c* towards the *ab* plane for both the zero-field-grown and field-grown $MnBi_2Te_4$ because the magnetic easy axis, in both cases, is along the *c* direction.

With obvious changes in the magnetic torque at $H_{SF}$, $H_1$, and $H_2$, it is truly surprising there is no manifestation in $MR_{ab}$, which varies linearly with H at $H > H_{SF}$ in all measured angles (Figure 3(d)). This observation suggests that there is little spin scattering in our field-grown $MnBi_2Te_4$. Similar linear magnetoresistance was also observed in zero-field-grown $MnBi_2Te_4$ films when $H > H_1$, which is attributed to the high surface conductivity [51]. On the other hand, the fact that $MR_{ab}$ monotonically decreases with increasing the angle suggests that the linear $MR_{ab}$ results from the unique electronic structure such as the existence of hot spots [52] or sharp corners [53] of the Fermi surface but unlikely from open orbits [54] as quantum oscillations are observed *vide infra*.

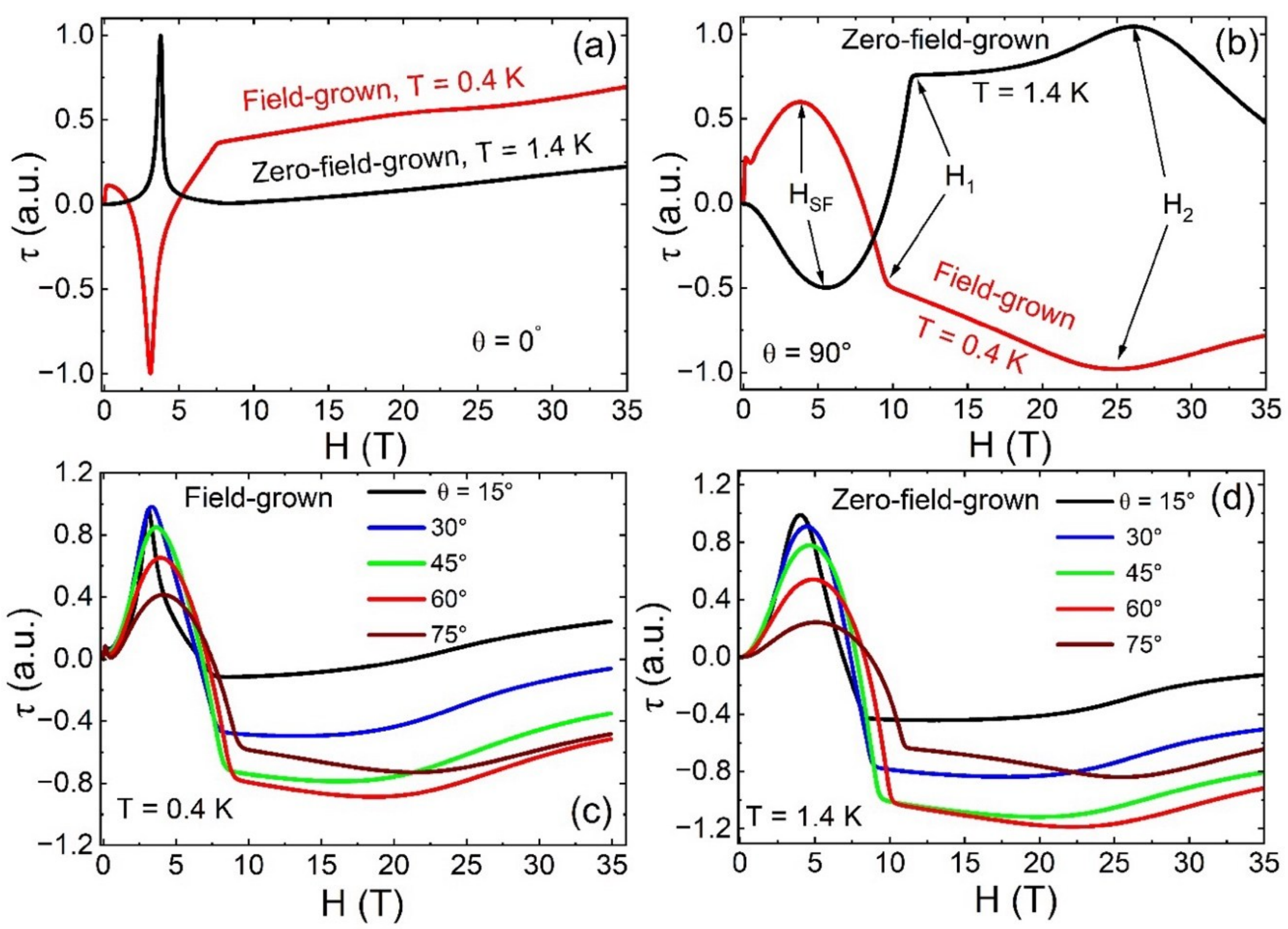

FIG. 4. (a-b) Field dependence of the magnetic torque for zero-field-grown and field-grown $MnBi_2Te_4$ with magnetic field applied along the *c* axis (θ = 0°) (a) and (b) *ab* plane (θ = 90°). (c-d) Field dependence of the magnetic torque for θ = 15°, 30°, 45°, 60°, 75° for (c) field-grown and (d) zero-field-grown $MnBi_2Te_4$.

For zero-field-grown $MnBi_2Te_4$, a topological phase transition is predicted to arise from a topological insulator in zero field to a type-II Weyl semimetal in the FM state [26,55]. Experimentally, the Shubnikov-de Haas (SdH) oscillation is observed above 10 T with the single frequency ~79.9 T for H // *c* [51]. Contrastingly for our field-grown $MnBi_2Te_4$, the field dependence of τ, with H // *c* at T = 0.4 K between 8 T and 35 T, possesses a weak oscillation, as shown in Figure 5(a). By subtracting the background represented by the red dashed line, Δτ is obtained which is plotted as a function of 1/H in Figure 5(b). Through a Fast Fourier Transformation (FFT) shown in Figure 5(c), the oscillation frequency F ≈ 40 T is obtained. In other words, the field-grown sample exhibits a de Haas-van Alphen (dHvA) oscillation with a frequency of about half the value reported for zero-field-grown $MnBi_2Te_4$ [51]. Figure 5(d) shows the Landau fan diagram constructed by assigning the oscillation maxima to n + 1/4 and minima to n – 1/4, where n is the Landau level index. The solid line in Figure 5(d) represents a linear fit of data with n = 0.3 + 39.7/H. According to the Lifshitz Onsager quantization criterion [54,56], the slope of the linear equation gives the oscillation frequency F = 39.7 T, consistent with the value obtained from the FFT spectrum. Since only one oscillation frequency is detected and the Hall resistivity is positive (Fig. 3(e)), this oscillation originates from a hole band.

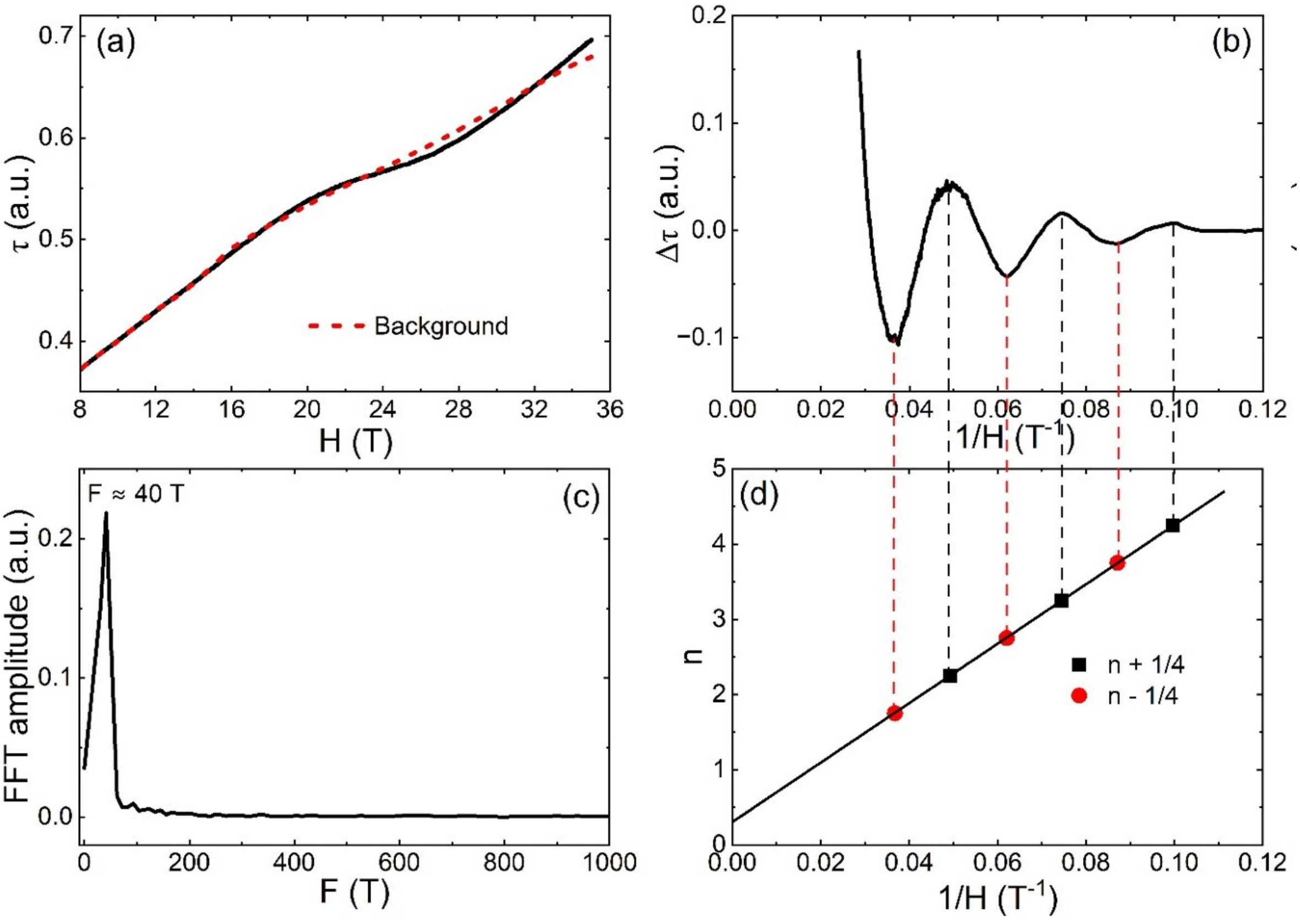

FIG. 5. (a) Magnetic field dependence of the magnetic torque (τ) of field-grown $MnBi_2Te_4$ at T = 0.4 K with H // *c* between 8 T and 35 T. The red curve represents the background $\tau_{bg}$. (b) Δτ versus 1/H. (c) FFT of Δτ revealing a peak at F ~ 40 T. (d) Landau fan diagram constructed from the dHvA oscillation.

Based on the above experimental observation, the electronic and magnetic properties of field-grown $MnBi_2Te_4$ are different from that of zero-field-grown case. Figure 6 presents the calculated band structure, by including spin-orbit coupling, for $MnBi_2Te_4$ with A-type AFM (a) and FM (b), respectively. Because of the AFM ordering, there is the band splitting in AFM $MnBi_2Te_4$, forming the insulating AFM state with the band gap 0.1390 eV. This result is consistent with earlier report for bulk $MnBi_2Te_4$ [26]. For $MnBi_2Te_4$ with the FM ground state (Figure 6(b)), the electronic structure shows mainly bands composed of Te with a small amount of Bi crossing the Fermi energy ($E_F$) (mixed green-orange color going from Γ to T in Figure 6(b)). This results in the metallic nature of FM $MnBi_2Te_4$ with a hole band located at T. The width of the hole band is 0.4256 $nm^{-1}$. If the above torque oscillation results from the hole cyclotron motion, we can experimentally estimate the band width from the frequency through the Onsager relation $F = (\Phi_0/2\pi^2)S_F$, where $\Phi_0$ is the magnetic flux quantum and $S_F$ represents the extremal Fermi surface cross-sectional area to the magnetic field. For F = 39.7 T, we obtain $S_F = 0.378$ $nm^{-2}$. With the approximation of a circular-shaped cross section, the corresponding Fermi wave vector is $k_F \sim 0.347$ $nm^{-1}$. This value is larger than the calculated hole band width mentioned above, which only used the crossing near the T point. Detailed Fermi surface topology is needed for understanding the difference between experiment and calculations.

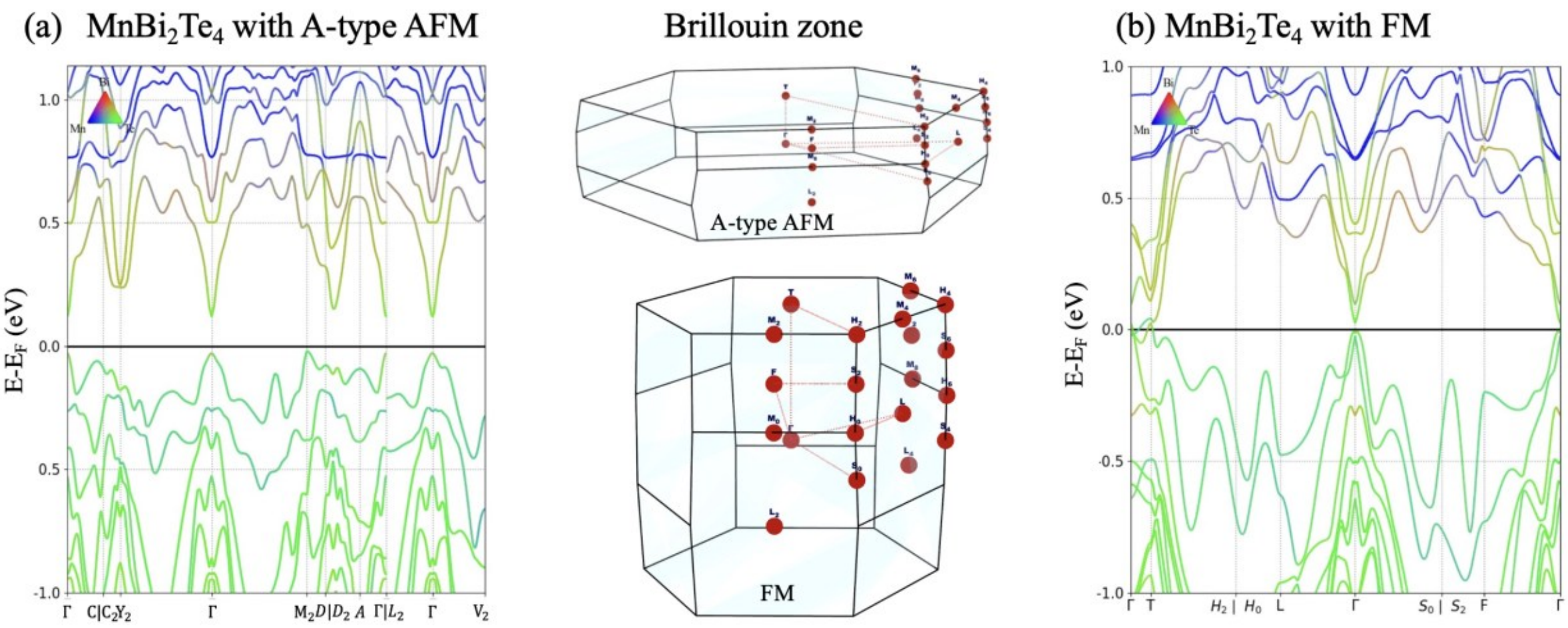

FIG. 6. Band structures for $MnBi_2Te_4$ with A-type AFM (a) and FM (b) ground states.

Interestingly, earlier calculations using (GGA) + U (U = 3 eV) functional with fixed lattice constants predict that FM $MnBi_2Te_4$ with the magnetic easy axis along the *c* direction is a type-II Weyl semimetal [55]. We recall the constant term of n versus 1/H relation is related to the Berry phase $\Phi_B$ with $(\Phi_B/2\pi) + \delta = 0.3$ for FM $MnBi_2Te_4$, where $\delta = 0$ for a 2D and $\pm$ 1/8 for a 3D Fermi surface (+ / - sign corresponds to the maxima/minima of the cross-sectional area of the Fermi surface for a hole band) [54,56]. Since only one oscillation frequency is detected and the Hall resistivity is positive (Fig. 3(e)), we consider that the oscillation is from a hole band with the minima of the cross-sectional area, i.e., $\delta = -1/8$. This leads to $\Phi_B \sim 0.85\pi$, pointing to a nontrivial band. If the probed Fermi surface has 2D characteristic with $\delta = 0$, $\Phi_B(2D) \sim 0.60\pi$.

According to calculations reported in Ref. [57], the electronic band structure in the vicinity of the Weyl points is highly sensitive to the orientation of the magnetic moments. Specifically, when the magnetic moment is aligned along the crystallographic *c* axis, the system exhibits type-II Weyl semimetal characteristics, where the Weyl cones are strongly tilted and electron and hole pockets coexist at the Fermi level. As the magnetic moment rotates away from the *c* axis and approaches alignment within the *ab* plane, the band structure gradually evolves toward a type-I Weyl phase, characterized by more symmetric, upright Weyl cones and well-defined point-like Fermi surfaces. The experimentally observed linear $MR_{ab}$ across all applied magnetic field directions, persisting up to 35 T as shown in Fig. 3(d), is consistent with the Weyl semimetal behavior [57] but not uniquely indicative as linear magnetoresistance can arise from other origins such as disorder or quantum-limit effects.

## 3. CONCLUSIONS

Based on theoretical predictions, the magnetic topological insulator $MnBi_2Te_4$ is expected to host multiple magnetic configurations in its metastable states. Here, we experimentally realize such a metastable state by synthesizing $MnBi_2Te_4$ single crystals under a 9 T magnetic field. Although the field-grown crystals retain the same rhombohedral crystal structure as their zero-field-grown counterparts, they display a distinctly different magnetic ground state: instead of the antiferromagnetic order characteristic of zero-field-grown $MnBi_2Te_4$, the field-grown crystals exhibit a ferromagnetic ground state with a Curie temperature of ~12.5 K. By measuring the magnetic torque in various field orientations, we observe features associated with spin reorientation, characterized by fields $H_{\mathrm{SF}} < H_1 < H_2$, in both zero-field-grown and field-grown crystals. The presence of these transitions in the field-grown samples indicates a compensated ferromagnetic configuration. Quantitative analysis of the low-temperature (0.4 K) torque data further reveals de Haas–van Alphen oscillations with a single frequency of 39.7 T. Together with the positive Hall resistivity, these results demonstrate that the Fermi surface of field-grown $MnBi_2Te_4$ differs significantly from that of the zero-field-grown $MnBi_2Te_4$, which is supported by DFT calculations. The experimentally probed hole band has a nontrivial Berry phase for H // $c$. Additionally, the magnetoresistance of the field-grown $MnBi_2Te_4$ exhibits a linear dependence on magnetic field and shows no anomalies across $H_{\mathrm{SF}}, H_1$, or $H_2$. This behavior suggests that the magnetoresistance arises predominantly from orbital contributions enhanced by strong spin–orbit coupling, rather than spin scattering.

Overall, our findings establish that magnetic-field-assisted synthesis provides an effective route to reconfigure the ground-state spin structure of $MnBi_2Te_4$. This capability to stabilize novel magnetic states offers a powerful means to tailor the electronic structure of the material. Demonstration of magneto-synthesis is also described in Ref. [58].

## 4. ACKNOWLEDGEMENTS

This project was supported by the grant No. DE-SC0024501 funded by the U. S. Department of Energy, Office of Science. The development of the high-field furnace insert and related instrumentation was supported by the U.S. Department of Energy's Office of Energy Efficiency and Renewable Energy (EERE) under the Industrial Efficiency & Decarbonization Office (IEDO) Award No. DE-EE0009131. The 9.4 T magnet and the associated facility were supported by the National High Magnetic Field Laboratory (NHMFL) funded by the National Science Foundation cooperative agreement DMR-1644779 and the State of Florida, with the other recent high magnetic field activities funded by DMR-2128556.

## 5. DATA AVAILABILITY

All the data supporting the findings in this study is available. Further data and methods are available from the corresponding author upon request.

## 6. AUTHOR CONTRIBUTION

A.R., M.W.M., and R.J. designed the experiments. A.R., A.M.D., J.M., M.W.M. and R.J. prepared and synthesized the samples. A.R. performed the experimental measurements up to 14 T. A.R., D.D., and D.G. ran high magnetic field measurements at NHMFL. G.M.Z., I.V., and W.A.S. conducted the DFT calculations. A.R., M.W.M., W.A.S., and R.J. drafted the manuscript. All authors contributed to manuscript preparation.

## 7. COMPETING INTEREST

The authors declare no competing interests.

## 8. REFERENCES

[1] Y. Tokura, K. Yasuda, and A. Tsukazaki, *Magnetic Topological Insulators*, Nature Reviews Physics **1**, 126 (2019).

[2] B. A. Bernevig, C. Felser, and H. Beidenkopf, *Progress and Prospects in Magnetic Topological Materials*, Nature **603**, 41 (2022).

[3] X. Zhang, X. Wang, T. He, L. Wang, W. W. Yu, Y. Liu, G. Liu, and Z. Cheng, *Magnetic Topological Materials in Two-Dimensional: Theory, Material Realization and Application Prospects*, Science Bulletin **68**, 2639 (2023).

[4] Y. Gong, J. Guo, J. Li, K. Zhu, M. Liao, X. Liu, Q. Zhang, L. Gu, L. Tang, X. Feng, D. Zhang, W. Li, C. Song et al., *Experimental Realization of an Intrinsic Magnetic Topological Insulator*, Chinese Physics Letters **36**, 076801 (2019).

[5] M. Z. Hasan and C. L. Kane, *Colloquium: Topological Insulators*, Rev. Mod. Phys. **82**, 3045 (2010).

[6] X. L. Qi and S. C. Zhang, *Topological Insulators and Superconductors*, Rev. Mod. Phys. **83**, 1057 (2011).

[7] C.-Z. Chang, J. Zhang, X. Feng, J. Shen Z. Zhang M. Guo, K. Li et al., *Experimental Observation of the Quantum Anomalous Hall Effect in a Magnetic Topological Insulator*, Science **340**, 167 (2013).

[8] W. Tian, W. Yu, J. Shi, and Y. Wang, *The Property, Preparation and Application of Topological Insulators: A Review*, Materials **10**, 814 (2017).

[9] M. M. Otrokov, I. I. Klimovskikh, H. Bentmann, D. Estyunin, A. Zeugner et al., *Prediction and Observation of an Antiferromagnetic Topological Insulator*, Nature **576**, 416 (2019).

[10] A. Zeugner, F. Nietschke, A. U. B. Wolter et al., *Chemical Aspects of the Candidate Antiferromagnetic Topological Insulator $MnBi_2Te_4$*, Chemistry of Materials **31**, 2795 (2019).

[11] J. Q. Yan, Q. Zhang, T. Heitmann, Z. Huang, K. Y. Chen, J. G. Cheng, W. Wu, D. Vaknin, B. C. Sales, and R. J. McQueeney, *Crystal Growth and Magnetic Structure of $MnBi_2Te_4$*, Phys. Rev. Mater. **3**, 064202 (2019).

[12] Y. Li, C. Liu, Y. Wang, Z. Lian, S. Li, H. Li, Y. Wu, H. Z. Lu, J. Zhang, and Y. Wang, *Giant Nonlocal Edge Conduction in the Axion Insulator State of $MnBi_2Te_4$*, Sci. Bull. (Beijing) **68**, 1252 (2023).

[13] D. A. Estyunin, I. I. Klimovskikh, A. M. Shikin et al., *Signatures of Temperature Driven Antiferromagnetic Transition in the Electronic Structure of Topological Insulator $MnBi_2Te_4$*, APL Mater. **8**, 021105 (2020).

[14] Y. F. Zhao et al., *Even-Odd Layer-Dependent Anomalous Hall Effect in Topological Magnet $MnBi_2Te_4$ Thin Films*, Nano Lett **21**, 7691 (2021).

[15] S. K. Bac, K. Koller, F. Lux et al., *Topological Response of the Anomalous Hall Effect in $MnBi_2Te_4$ due to Magnetic Canting*, NPJ Quantum Mater. **7**, 46 (2022).

[16] J. Luo et al., *Exploring the Epitaxial Growth Kinetics and Anomalous Hall Effect in Magnetic Topological Insulator $MnBi_2Te_4$ Films*, ACS Nano **17**, 19022 (2023).

[17] A. R. Mazza et al., *Surface-Driven Evolution of the Anomalous Hall Effect in Magnetic Topological Insulator $MnBi_2Te_4$ Thin Films*, Adv. Funct. Mater. **32**, 2202234 (2022).

[18] K. Zhu, Y. Cheng, M. Liao, S. K. Chong, D. Zhang, K. He, K. L. Wang, K. Chang, and P. Deng, *Unveiling the Anomalous Hall Response of the Magnetic Structure Changes in the Epitaxial $MnBi_2Te_4$ Films*, Nano Lett **24**, 2181 (2024).

[19] J. Cui, M. Shi, H. Wang, F. Yu, T. Wu, X. Luo, J. Ying, and X. Chen, *Transport Properties of Thin Flakes of the Antiferromagnetic Topological Insulator $MnBi_2Te_4$*, Phys. Rev. B **99**, 155125 (2019).

[20] J. Q. Yan, S. Okamoto, M. A. McGuire, A. F. May, R. J. McQueeney, and B. C. Sales, *Evolution of Structural, Magnetic, and Transport Properties in* $MnBi_{2-x}Sb_xTe_4$, Phys. Rev. B **100**, 104409 (2019).

[21] I. I. Klimovskikh, M. M. Otrokov, D. Estyunin et al., *Tunable 3D/2D Magnetism in the* $(MnBi_2Te_4)(Bi_2Te_3)_m$ *Topological Insulators Family*, NPJ Quantum Mater. **5**, 54 (2020).

[22] S. Changdar, S. Ghosh, K. Vijay, I. Kar, S. Routh, P. K. Maheshwari, S. Ghorai, S. Banik, and S. Thirupathaiah, *Nonmagnetic Sn Doping Effect on the Electronic and Magnetic Properties of Antiferromagnetic Topological Insulator* $MnBi_2Te_4$, Physica B **657**, 414799 (2023).

[23] S. Yang and Y. Ye, *Intrinsic and Defect-Related Magnetism of* $MnBi_2Te_4(Bi_2Te_3)_n$ *Family from the Bulk to Two-Dimensional Limit*, 2D Materials **12**, 012003 (2024).

[24] Y. J. Hao, P. Liu, Y. Feng, X. Ma, E. F. Schwier et al., *Gapless Surface Dirac Cone in Antiferromagnetic Topological Insulator* $MnBi_2Te_4$, Phys. Rev. X **91**, 041038 (2019).

[25] C. Hu, K. N. Gordon, P. Liu, J. Liu, X. Zhou et al., *A van der Waals Antiferromagnetic Topological Insulator with Weak Interlayer Magnetic Coupling*, Nat. Commun. **11**, (2020).

[26] J. Li, Y. Li, S. Du, Z. Wang, B.-L. Gu, S.-C. Zhang, K. He, W. Duan, and Y. Xu, *Intrinsic Magnetic Topological Insulators in van der Waals Layered* $MnBi_2Te_4$*-Family Materials*, Sci. Adv. **12**, aaw5685 (2019).

[27] M. Z. Shi, B. Lei, C. S. Zhu, D. H. Ma, J. H. Cui, Z. L. Sun, J. J. Ying, and X. H. Chen, *Magnetic and Transport Properties in the Magnetic Topological Insulators* $MnBi_2Te_4(Bi_2Te_3)_n$ $(n = 1, 2)$, Phys Rev B **100**, 155144 (2019).

[28] A. Rajbanshi, J. Xing, D. Gong, W. A. Shelton, and R. Jin, *An Alternative Ground State of* $MnBi_2Te_4$ *Obtained by Magnetic Annealing*, APL Mater. **13**, 101110 (2025).

[29] Y. Du, H. Zhang, F. Zhou, T. Wang, J. Li, W. Qi, Y. Zhang, Y. Yu, F. Fei, and F. Song, *Magnetic Properties Manipulation in* $MnBi_2Te_4$ *through Electrochemical Organic Molecular Intercalation*, Chinese Physics B **34**, 087302 (2025).

[30] M. H. Du, J. Yan, V. R. Cooper, and M. Eisenbach, *Tuning Fermi Levels in Intrinsic Antiferromagnetic Topological Insulators* $MnBi_2Te_4$ *and* $MnBi_4Te_7$ *by Defect Engineering and Chemical Doping*, Adv. Funct. Mater. **31**, 2006516 (2021).

[31] C. Liu, Y. Wang, M. Yang, J. Mao, H. Li, Y. Li, J. Li, H. Zhu, J. Wang, L. Li, Y. Wu, Y. Xu, J. Zhang, Y. Wang, *Magnetic-field-induced Robust Zero Hall Plateau State in* $MnBi_2Te_4$ *Chern Insulator*, Nat. Commun. **12**, 4647 (2021).

[32] C. Pei, Y. Xia, J. Wu, Y. Zhao, L. Gao, T. Ying, B. Gao, N. Li, W. Yang, D. Zhang, H. Gou, Y. Chen, H. Hosono, G. Li, Y. Qi, *Pressure-induced Topological and Structural Phase Transitions in an Antiferromagnetic Topological* Insulator, Chin. Phys. Lett. **37**, 066401 (2020).

[33] W. Guo, L. Huang, Y. Yang, Z. Huang, J. Zhang, *Pressure-induced Topological Quantum Phase Transition in the Magnetic Topological Insulator $MnBi_2Te_4$*, New J. Phys. **23**, 083030 (2021).

[34] Z. Xu, M. Ye, J. Li, W. Duan, Y. Xu, *Hydrostatic Pressure-induced Magnetic and Topological Phase Transitions in the $MnBi_2Te_4$ Family of Materials*, Phys. Rev. B **105**, 085129 (2022).

[35] A. Yu. Vyazovskaya, M. Bosnar, E. V. Chulkov, M. M. Otrokov, *Intrinsic Magnetic Topological Insulators of the $MnBi_2Te_4$ Family*, Commun. Mat. **6**, 88 (2025).

[36] Q. Jiang, C. Wang, P. Malinowski, Z. Liu, Y. Shi, Z. Lin, Z. Fei, T. Song, D. Graf, X. Xu, J. Yan, D. Xiao, J. W. Chu, Phys. Rev. B **103**, 205111 (2021).

[37] S. Flynn, C. L. Benyacko, M. Mihalik, J. Lee, F. Ma et al., *Synthesis of Cobalt Grown from Co-S Eutectic in High Magnetic* Fields, Phys. Rev. Mater. **9**, 094401(2025).

[38] G. Kresse and J. Furthmüller, *Efficient Iterative Schemes for ab initio Total-energy Calculations Using a Plane-wave Basis Set*, Phys. Rev. B **54**, 11169 (1996).

[39] G. Kresse and J. Hafner, Ab initio Molecular Dynamics for Liquid Metals, Phys. Rev. B **47**, 558 (1993).

[40] G. Kresse and D. Joubert, *From Ultrasoft Pseudopotentials to the Projector Augmented-wave Method*, Phys. Rev. B **59**, 1758 (1999).

[41] P. E. Blöchl, *Projector Augmented-wave Method*, Phys. Rev. B **50**, 17953 (1994).

[42] J. Klimeš, D. R. Bowler and A. Michaelides, *Chemical Accuracy for the van der Waals Density Functional*, J. Phys.: Condens. Matter **22**, 022201 (2010).

[43] L. Ding, C. Hu, F. Ye, E. Feng, N. Ni, H. Cao, *Crystal and Magnetic Structures of Magnetic Topological Insulators $MnBi_2Te_4$ and $MnBi_4Te_7$*, Phys. Rev. B **101**, 020412(R) (2020).

[44] Y. Lai, L. Ke, J. Yan, R. D. McDonald, R. J. McQueeney, *Defect-driven Ferrimagnetism and Hidden Magnetization in $MnBi_2Te_4$*, Phys. Rev. B **103**, 184429 (2021).

[45] T. Das, S. Mukhopadhyay, *Metamagnetic Quantum Criticality in the Antiferromagnetic Topological Insulator $MnBi_2Te_4$*, Phys. Rev. B **111**, 174420 (2025).

[46] R. A. Susilo, C. H. Hsu, H. Lin, J. M. Cadogan, W. D. Hutchison, S. J. Campbell, *Structural, Thermal and Magnetic Properties of $Y_2Fe_2Si_2C$,* J. Alloys & Compounds **778**, 618 (2019).

[47] T. Murakami, Y. Nambu, T. Koretsune, X. Gu, T. Yamamoto, C. M. Brown, H. Kageyama, *Realization of Interlayer Ferromagnetic Interaction in $MnSb_2Te_4$ toward the Magnetic Weyl Semimetal State*, Phys. Rev. B **100**, 195103 (2019).

[48] X. Lei, L. Zhou, Z. Y. Hao, X. Z. Ma, C. Ma, Y. Q. Wang, P. B. Chen, B. C. Ye, L. Wang, F. Ye, J. N. Wang, J. W. Mei, H. T. He, *Surface-induced Linear Magnetoresistance in the Antiferromagnetic Topological Insulator $MnBi_2Te_4$*, Phys. Rev. B **102**, 235431 (2020).

[49] Z. Yang, H. Zhang, *Evolution of Surface States of Antiferromagnetic Topological Insulator $MnBi_2Te_4$ with Tuning the Surface Magnetization*, New J. Phys. **24**, 073034 (2022).

[50] S. H. Lee, Y. Zhu, Y. Wang, L. Miao, T. Pillsbury, H. Yi, S. Kempinger, J. Hu, C. A. Heikes, P. Quarterman, W. Ratcliff, J. A. Borchers, H. Zhang, X. Ke, D. Graf, N. Alem, C. Z. Chang, N. Samarth, Z. Mao, *Spin Scattering and Noncollinear Spin Structure-induced Intrinsic Anomalous Hall Effect in Antiferromagnetic Topological Insulator $MnBi_2Te_4$*, Phys. Rev. Res. **1**, 012011(R) (2019).

[51] X. Lei, L. Zhou, Z. Y. Hao, H. T. Liu, S. Yang, H. P. Sun, X. Z. Ma, C. Ma, H. Z. Lu, J. W. Mei, J. N. Wang, H. T. He, *Magnetically Tunable Shubnikov-de Haas oscillations in* $MnBi_2Te_4$, Phys. Rev. B **105**, 155402 (2022).

[52] N. Maksimovic, I. M. Hayes, V. Nagarajan, J. G. Analytis, A. E. Koshelev, J. Singleton, Y. Lee, T. Schenkel, *Magnetoresistance Scaling and the Origin of H-Linear Resistivity in $BaFe_2(As_{1-x}P_x)_2$*, Phys. Rev. X **10**, 041062 (2020).

[53] R. D. H. Hinlopen, F. A. Hinlopen, J. Ayres, N. E. Hussey, *$B^2$ to B-linear Magnetoresistance due to Impeded Orbital Motion*, Phys. Rev. Res. **4**, 033195 (2022).

[54] D. Shoenberg, *Magnetic Oscillations in Metals* (Cambridge University Press, 1984).

[55] D. Zhang, M. Shi, T. Zhu, D. Xing, H. Zhang, J. Wang, *Topological Axion States in the Magnetic Insulator $MnBi_2Te_4$ with the Quantized Magnetoelectric Effect*, Phys. Rev. Lett. **122**, 206401 (2019).

[56] R. Chapai, P. V. S. Reddy, L. Xing, D. E. Graf, A. B. Karki, T. R. Chang, and R. Jin, *Evidence for Unconventional Superconductivity and Nontrivial Topology in PdTe*, Sci. Rep. **13**, 6824 (2023).

[57] J. Li, C. Wang, Z. Zhang, B. Gu, W. Duan, Y. Xu, *Magnetically Controllable Topological Quantum Phase Transitions in the Antiferromagnetic Topological Insulator $MnBi_2Te_4$*, Phys. Rev. B **100**, 121103(R) (2019).

[58] Gang Cao, Lance DeLong, *Physics of Spin-Orbit-Couples Oxides* (Oxford University Press).